\documentstyle[pre,aps,psfig]{revtex}
\newcommand{\be}{\begin{eqnarray}}
\newcommand{\ee}{\end{eqnarray}}
\newcommand{\Herm}{\mbox{H}}
\newcommand{\im}{\mbox{Im}}
\newcommand{\re}{\mbox{Re}}
\newcommand{\kBT}{k_{B} T}
\newcommand{\nxyz}{n_{x}n_{y}n_{z}}
\newcommand{\E}{\cdot10^}
\newcommand{\Tl}{T_{\parallel}}
\newcommand{\Tt}{T_{\perp}}
\newcommand{\dd}{\mbox{d}}

\begin{document}
\twocolumn[\hsize\textwidth\columnwidth\hsize
           \csname @twocolumnfalse\endcsname
\title{Dielectric properties of interacting storage ring plasmas}

\author{A. Selchow, K. Morawetz}
\address{Fachbereich Physik, Universit\"at Rostock, D-18051 Rostock, Germany}
\date{\today}
\maketitle

\begin{abstract}
A dielectric function  including collisional correlations is derived by linearizing 
the self consistent Vlasov equation with a
Fokker-Planck collision integral.
The calculation yields the same type of dielectric function  as in the standard-theory of
Schottky-noise in storage rings. This dielectric function  is compared with the Mermin-dielectric function 
derived from a kinetic equation with relaxation time approximation. We observe
that these functions are identical, however the Mermin-DF is computationally
advantageous.
The limits of both dielectric functions are given and the sum rules are
proven. We apply these dielectric functions for typical storage
ring plasmas and calculate the stopping power and the plasmon
excitation spectrum.
\end{abstract}
\pacs{PACS Numbers: ??}
\vskip2pc]

\section{Introduction}

During the last ten years experiments with ions ( p$^+ $ up to U$^{92+}$) 
in storage rings gained importance in the field of spectroscopy and
plasma physics.  The storaged and cooled ion beams have a high luminosity for 
recombination experiments and inertial confined fusion investigations. In 
particular it is of basic interest to study the transition between the 
weak and strong coupled plasma or even the transition to a crystalline state 
of a cooled ion beam \cite{Habs95}. The most important prerequisite for 
obtaining
dense states is strong electron and laser cooling. The electron cooling 
force can be described as stopping power acting on an ion beam in an electron 
plasma \cite{Poth90}. Another plasma phenomena in dense beams are
collective excitations (plasmons, shear modes) which are detectable
by the Schottky noise \cite{Park80}. All items - the 
pair distribution function of a state, the stopping power and the shape of
the collective excitations are related to the dielectric function 
$ \epsilon(\vec q, \omega) $.

Within the linear response theory the polarizability $ \Pi(\vec q, \omega)$
(and altogether the dielectric function  $ \epsilon(\vec q, \omega) $) is defined by the
variation of particle density $\delta n(\vec q,\omega)$ in reaction to an
external field $ \delta U_{ext}(\vec q, \omega) $ via
\be
\label{PolariDef}
\Pi(\vec q,\omega) &=& \frac{\delta n(\vec q,\omega)}{\delta U_{ext}} \; .
\ee
The connection to the dielectric function (DF) is given by
\be
\label{epsilonDef}
\epsilon(\vec q, \omega) &=& 1 + V_C \Pi(\vec q, \omega) \; .
\ee

The captured ions in storage rings are moving in front of a background
of the confining fields ensuring approximately the charge neutrality in
the system. This nearly neutral system of ions interacting via the
Coulomb potential $ V_C $ immersed in a homogeneous background of 
opposite charge is usually called an one-component-plasma (OCP). 

An unpleasant problem is the temperature anisotropy. The longitudinal 
temperature ($\Tl$) differs from the transversal ($\Tt$) (referring to the 
beam-axis) because only the longitudinal direction is cooled directly. 
Even taking into account relaxation processes between the degrees of freedom
and possible transversal cooling the temperature difference maintains.

In this paper we focus on experiments done by the Heidelberg TSR group
with  $ ^9$Be$^+$ and $^{12}$C$^{6+}$ ions cooled 
by an electron beam \cite{Habs95,Sram97}.
The Be-ions can be cooled further down to few mK by applying laser cooling.

For estimations about the 
ideality and collision-numbers we employ the longitudinal temperature $\Tl$.
The density of the ion beam can be calculated using the current $j$ of the 
beam 
profile (the diameter) measured by the beam profile monitors $ x_{BPM} $ and 
the value of the betatron function on this position $ \beta_{BPM} $
\be
n = \frac{ j Q \beta_{BPM} }{2\pi Z x^2_{BPM} R v_0} \,.
\ee
Here denotes $ v_0 $ the ion beam velocity, $R$ is the ring radius and $Z$ is
the charge state of the ions. The transversal tune $ Q $ amounts to 2.8.
\begin{table}[bt]
\begin{tabular}{|l|l|l|l|}
\hline
  Parameter &  Be$^+$         & C$^{6+}$           &  $e^-$              \\
\hline
  $n$       & $ 2.3\E{13}\mbox{m}^{-3} $ & $ 1.6\E{13}\mbox{m}^{-3}$ 
                      & $2.9\E{13}\mbox{m}^{-3}$                         \\
\hline
$ \Tl $     & 6K              & $8000$K            & $ 3.5$K             \\ 
\hline
$  \Gamma $ & $ 0.13 $        & $ 0.0031$          & $ 0.23 $            \\
\hline
$ \lambda $ &  $ 1.7\E{7}\mbox{s}^{-1} $ & $ 4.0\E{5}\mbox{s}^{-1} $ 
       &  $ 3.4\E{12}\mbox{s}^{-1}$ \\
\hline
$ \omega_{pl} $  & $ 2.1\E{6}\mbox{s}^{-1} $ & $ 9.2\E{6}\mbox{s}^{-1} $  
    & $ 3.0\E{8}\mbox{s}^{-1} $ \\
\hline
\end{tabular}
\caption{\label{ParaTab} Parameters for several experiments in the TSR. The explanations
          are given in the text. Data are from \protect\cite{Habs95,Sram97}. }
\end{table}
The essential parameter for characterizing the coupling is the
nonideality or plasma parameter
\be
  \Gamma = \frac{e^{2}}{ 4\pi\epsilon_{0} k_{B}T }
  \left(\frac{4\pi n}{3}\right)^{\frac{1}{3}}
\ee
which is the ratio of potential and thermal energy. 
Further essential quantities are the plasma-frequency
\be
\omega_{pl} = \sqrt{\frac{ne^2}{\epsilon_0 m}}
\ee
and the inverse screening length
\be
\label{wavenumber}
\kappa = \sqrt{\frac{ne^2}{\epsilon_0 \kBT}} \;.
\ee
Another important parameter characterizing the plasma is the collision
frequency, or friction coefficient $\lambda$ which is the inverse
relaxation time
\be
\lambda = \frac{1}{\tau} = n \sigma(v_{th}) v_{th} \\
\nonumber
\sigma(v) = 4\pi \left(\frac{2 e^2}{12 \epsilon_{0} k_{B} T}\right) \Lambda(v).
\ee
Here is $\sigma(v_{th})$ the cross-section at thermal velocity 
$v_{th}^2=2 k_B T/m$ for ion-ion 
collisions, $ \Lambda(v) $ denotes the Coulomb logarithm e.g. in Brooks-Herring
approximation. 

The collisions between the ions play an essential role in this
storage-ring-plasmas being responsible for effects like intra-beam scattering
(IBS) \cite{Sore87}. That means an expansion of the (ion) beam due to 
ion-ion collisions has to be suppressed by electron cooling. Consequently for a 
sufficient description a dielectric function  $ \epsilon(\vec q, \omega)$ including
these collisions should be considered \cite{RW98}. In this paper two practical ways 
will be shown in chapter \ref{DF+Coll}.

An essential property of every dielectric function  is the fulfillment of the sum rules.
The most strongest are the longitudinal frequency sum rule
\be
\label{fsumrule}
\int_0^{\infty} \frac{2 \omega}{\pi \omega_{pl}^2}
\im \epsilon^{-1}(\vec q,\omega) \dd\omega  &=& 1
\ee
and the conductivity sum rule
\be
\label{ssumrule}
\int_0^{\infty} \frac{2 \omega}{\pi\omega_{pl}^2 }
\mbox{Im}\epsilon(\vec q,\omega) \dd\omega &=& -1  \;,
\ee
moreover the compressibility sum rule
\be
\label{f0sumrule}
\lim\limits_{q \to 0} \int_0^{\infty} 
\frac{2}{\pi \omega_{pl}^2}
{1 \over \omega} \im \epsilon^{-1}(\vec q,\omega) \dd\omega  &=& 1
\ee
and the perfect screening sum rule
\be
\label{s0sumrule}
\lim\limits_{q \to 0} \int_0^{\infty} 
\frac{2}{\pi\omega_{pl}^2 }
{1 \over \omega} \mbox{Im}\epsilon(\vec q,\omega) \dd\omega &=& -1.
\ee
The validity of these sum rules is an essential statement about the quality and
reliability of the dielectric function.

The outline of the paper is as follows: In Chapter \ref{MerminDF_ch} we give a
short rederivation of Mermin DF and the DF from a Fokker-Planck equation is given in Chapter \ref{VFPDF_ch}. In Chapter 
\ref{comp} we compare both DF's and the 
sum rules are proven in Chapter \ref{sumrules}. In Chapter \ref{SP} we present the application to the stopping power and 
in Chapter \ref{Schottky}
we calculate  the Schottky spectra.

\section{Dielectric functions with correlations}
\label{DF+Coll}

\subsection{Mermin-dielectric function }
\label{MerminDF_ch}

For calculating a dielectric function  including collisions between the 
particles with mass $m$, N. D. Mermin \cite{Merm70} suggested a particle 
number conserving dielectric function. We shortly sketch his derivation for 
the classical case starting with a kinetic equation in relaxation time 
approximation
\be
\label{Vlassov}
&&\frac{\partial}{\partial t} f(\vec r, \vec v, t) +
\vec v \frac{\partial}{\partial \vec r}  f(\vec r, \vec v, t)
+ \frac{\partial}{\partial \vec r} \frac{U(\vec r,t)}{m}
\frac{\partial}{\partial \vec v} f(\vec r, \vec v, t)
=\nonumber\\
&& -\frac{f(\vec r, \vec v, t)-f_0(\vec v)}{\tau} \;.
\ee
This kinetic equation describes the development
of a particle distribution function $ f(\vec r, \vec v, t) $ consisting of
an equilibrium part $ f_0(\vec v ) $ and a non-equilibrium part
$ \delta f(\vec r, \vec v,t)$
\be
 f(\vec r, \vec v, t) = f_0(\vec v ) + \delta f(\vec r, \vec v,t) \;.
\ee
The mean field $ U(\vec r,t)$ is composed of an external part and a part
arising from the induced particle density $\delta n$
\be
U(\vec{q,\omega}) = \delta U_{ext} + V \delta n(\vec{q},\omega) \;.
\ee
One gets the induced particle density $\delta n$ by linearization
of (\ref{Vlassov}) and integrating the solution
of $\delta f$ over the velocity $\vec v$. After Fourier
transformation $t\to \omega$ and $r \to q$
the following polarization function $ \Pi(\vec q,\omega) $ is obtained
\be
\label{deln-Pi}
\delta\! n(\vec{q},\omega) &=& \int \delta\!f(\vec q, \vec v,\omega) \dd^3 v
\nonumber\\
&=& \frac{\Pi_0(\vec q,\omega+{i \over \tau})}
{1 - V(\vec{q}) \Pi_0(\vec{q},\omega+{i \over \tau})} 
\delta U_{ext}(\vec{q},\omega)
\ee
with the RPA or Lindhard polarization function
\be
\label{Lindhard}
\Pi_0(\vec q, \omega) = \int \dd^3 v \frac{
\frac{\vec q}{m} 
\frac{\partial}{\partial \vec v} f(\vec v)} 
{ \vec v \vec q - \omega + i/\tau} \,.
\ee
The RPA dielectric function (\ref{epsilonDef}) in classical limit reads 
\be
\label{RPADF}
\epsilon(q,\omega) &=& 1 + \frac{\kappa^2}{q^2}  
 \bigg( 1 - 2x_c {\rm e}^{-x_c^{2}}
  \int_{0}^{x_c} {\rm e}^{t^{2}} \dd t
  + i \sqrt{\pi} x_c {\rm e}^{-x_c^{2}} 
\bigg)  
\nonumber\\
x_c &=& \sqrt{\frac{m}{2k_B T}} \frac{\omega}{q}
\ee
and fulfills all sum rules (\ref{fsumrule})-(\ref{s0sumrule}). Shifting the frequency into the complex plane according to \ref{Lindhard} 
one gets the relaxation
dielectric function. This expression does not fulfill the limit of static
screening and has a non Drude-like high frequency behavior which leads to a violation
of the sum rules (\ref{fsumrule})-(\ref{s0sumrule}).

In \cite{Merm70} was suggested a more sophisticated dielectric function by
considering the relaxation ansatz 
\be
\label{Vlassov1}
&&\frac{\partial}{\partial t} f(\vec r, \vec v, t) +
\vec v \frac{\partial}{\partial \vec r}  f(\vec r, \vec v, t)
+ \frac{\partial}{\partial \vec r} \frac{U(\vec r,t)}{m}
\frac{\partial}{\partial \vec v} f(\vec r, \vec v, t)
= \nonumber\\
&&-\frac{f(\vec r, \vec v, t)-\tilde f_0(\vec r,\vec v,t)}{\tau} \;.
\ee
with respect to a local equilibrium distribution function 
\be
\tilde {f_0}(\vec r,\vec v,t) = \exp\left[-\frac{m v^2}{2 \kBT} +
\frac{\mu +\delta\mu(\vec r,t)}{\kBT}\right]\nonumber\\
\ee
instead of the global distribution $f_0(\vec v)$ in (\ref{Vlassov}). In the simplest case one
can specify the local distribution by a small fluctuation in the chemical 
potential $\delta \mu$ related to the density-fluctuation $ \delta n$.

The Mermin-dielectric function is derived by solving 
(\ref{Vlassov1}) using an expansion of the local equilibrium distribution 
function in powers of $ \delta \mu $
\be
\label{collTerm}
\tilde f_0(\vec q,\vec v,\omega) = f_0(\vec v)-{\vec q \vec{\partial_v} 
f_0(\vec v)
\over m \vec q \vec v}\delta \mu(\vec q,\omega).
\ee
$ \delta \mu $ is determined by the particle number conservation 
$\omega \delta n(\vec q,\omega) = \int {\vec v \vec q} \delta f(\vec q,
\vec v, \omega) \dd^3 v $.
leading to \cite{Merm70}
\be
\delta \mu(\vec q,\omega) = \frac{\delta n(\vec q,\omega)}{\Pi(\vec q,0)} \;.
\ee
Finally one obtains from (\ref{Vlassov1}) and (\ref{collTerm}) for the
polarization function
\be
\Pi_{\rm M}(\vec q,\omega) = \frac{\Pi_0(\vec q,\omega+\frac{i}{\tau}) }
{\displaystyle 1 - \frac{1}{1-i\omega\tau} \left(1 -
\frac{\Pi_0(\vec q,\omega+\frac{i}{\tau})}{\Pi_0(\vec q,0)}
\right)}\;.
\ee
Instead of (\ref{deln-Pi}) we arrive at a density variation
\be
\label{deln-Pi1}
\delta\! n(\vec{q},\omega)
= \frac{\Pi_{\rm M}(\vec{q},\omega+{i \over \tau})}
{1 - V(\vec{q}) \Pi_{\rm M}(\vec{q},\omega+{i \over \tau})} \delta
U_{ext}(\vec{q},\omega)
\ee
and with (\ref{epsilonDef}) the Mermin-dielectric function 
finally has the shape
\be
\label{MerminDF}
\epsilon_{\rm M} \left( \vec q,\omega+\frac{i}{\tau} \right) = 1 + \frac{
  \left(1+ \frac{i}{\omega\tau}\right)
  (\epsilon(\vec q,\omega+\frac{i}{\tau}) - 1) }
{ \displaystyle
1 + \frac{i}{\omega\tau} \frac{ \epsilon(\vec q,\omega+\frac{i}{\tau}) - 1}
                                { \epsilon(\vec q,0) - 1}} \;.
\ee
Here denotes $ \epsilon(\vec q,\omega+\frac{i}{\tau}) $ the 
dielectric function (\ref{RPADF}) in relaxation time approximation. It is easy to see, that 
in the limit $ \tau \rightarrow \infty $ the Mermin-dielectric function  
reproduces the RPA-dielectric function (\ref{RPADF}).

\subsection{The Vlasov-Fokker-Planck Equation}
\label{VFPDF_ch}

Now we examine another kinetic equation - the Vlasov equation with the 
Fokker-Planck collision integral which has been used to predict the Schottky 
noise of an ion beam \cite{Park80}
\be
\label{VFPE}
&&\frac{\partial}{\partial t} f(\vec r, \vec v, t) +
\vec v \frac{\partial}{\partial \vec r}  f(\vec r, \vec v, t)
+ \frac{\partial}{\partial \vec r} \frac{U(\vec r,t)}{m}
\frac{\partial}{\partial \vec v}
f(\vec r, \vec v, t)\nonumber\\
&&= \lambda \frac{\partial}{\partial \vec v} \left( \frac{D}{\lambda}
\frac{\partial}{\partial \vec v} + \vec v \right) f(\vec r, \vec v, t)  \;.
\ee
The application of the Fokker-Planck collision term is valid for weak
collisions (it means low $q$-values) because it represents an expansion of the
collision integral in momentum space. With the collision integral of
the Fokker-Planck equation one includes the fluctuations of the distribution
function due to collisions. It describes the balance between dynamical
friction
$
\lambda \frac{\partial}{\partial \vec v}
\left( \vec v f(\vec r, \vec v,t) \right)
$
holding the velocity-distribution sharply near zero-velocity
and the diffusion 
$
D \frac {\partial^2}{\partial v^2} f(\vec r,\vec v,t)
$
flattening the velocity distribution. The coefficients $\lambda$ and $D$ in
the Fokker-Planck equation are related by the Einstein relation
\be
\frac{D}{\lambda} = \frac{\kBT}{m} \;.
\ee
As already mentioned above the friction coefficient $\lambda $ is
equal to the inverse relaxation time. Obviously,
the drift coefficient $\lambda \vec v$ is linear in the velocity as long as the diffusion 
coefficient $D$ is a constant. The Fokker-Planck collision term ensures the
particle conservation. Due to the Einstein relation we have a
proper balance between friction and diffusion. So we expect that
similar physics is included as the Mermin extension of the
simple relaxation time approximation in the last paragraph.

We solve this Fokker-Planck equation again within the linear response.
A sketch of the derivation can be found in appendix A with the
result for the dielectric function 
\be
&&\epsilon_{\rm VFP}(q,\omega) = 1 + \frac{\kappa^2}{q^2}\nonumber\\
&& \times\left( 1 +
\frac{i \omega}{\frac{\kBT}{m}\frac{q^2}{\lambda} - i \omega}
\left. _1F_1\right.
\left[1, 1 + \frac{\kBT}{m\lambda^2}q^2 - i\frac{\omega}{\lambda};
\frac{\kBT}{m\lambda^2} q^2 \right] \right) \;\nonumber\\
&&
\ee
and $_1F_1$ denotes the confluent hypergeometric function. This dielectric 
function has been given in \cite{Park80} and is valid for an isotropic 
plasma in three dimensions.

\onecolumn
\widetext
\begin{figure}[tb]
\psfig{figure=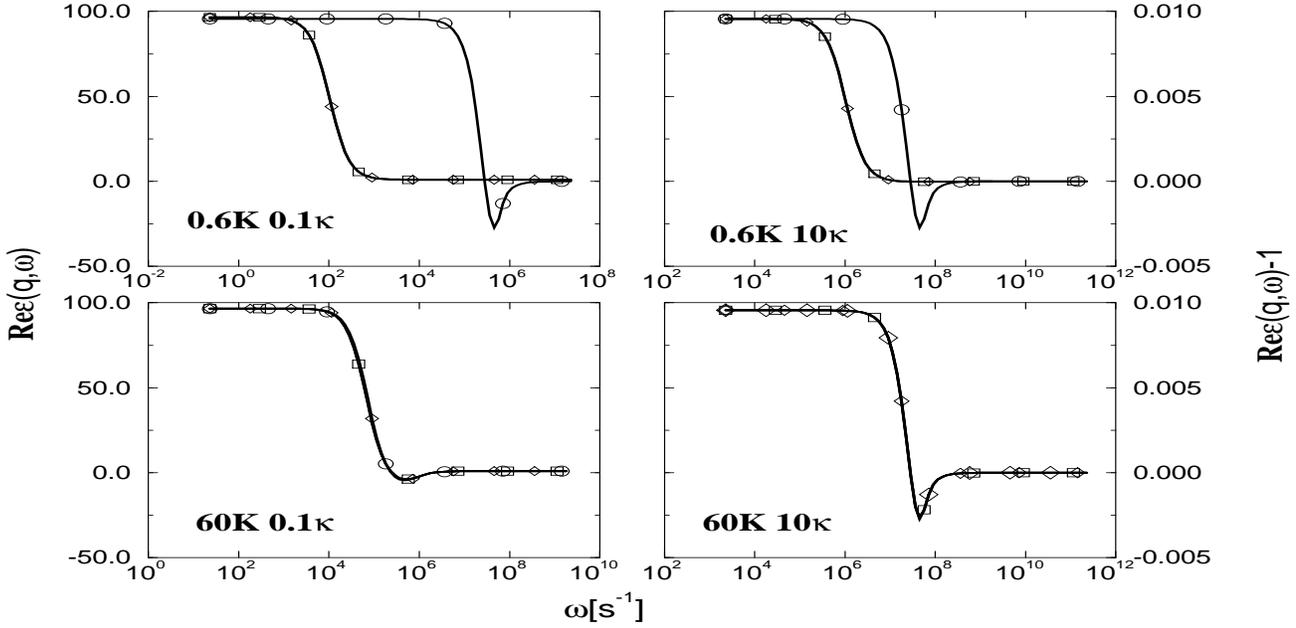,angle=-90,width=16cm,height=9cm,rheight=10cm}
\caption{\label{compare}
         Comparison of the real parts of the RPA- (circles), Mermin- (squares) 
         and VFP-DF 
         (diamonds). We have chosen temperatures of $ 0.6 \cdots 60 $K 
         available in the longitudinal direction of an ion beam and wave 
         numbers below and above the inverse Debye length
         $\kappa$ (\ref{wavenumber}). The particle density is
         $ n = 2.3\E{13}\mbox{m}^{-3}$ of single charged Beryllium ions.
         The real parts of Mermin- and the VFP-DF are identical.} 
\end{figure}
\begin{figure}[tb]
\psfig{figure=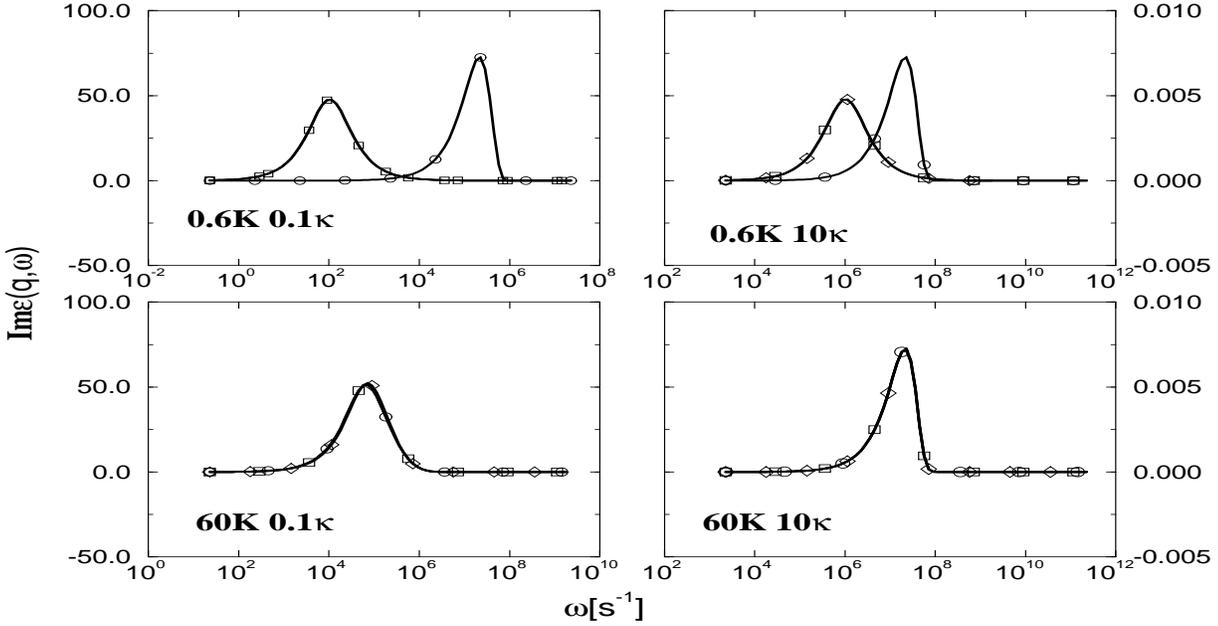,angle=-90,width=16cm,height=9cm,rheight=10cm
}
\caption{\label{compare1} Comparison of the imaginary parts of the RPA- (circles), Mermin- 
(squares) and VFP-DF (diamonds). The same density and temperatures are chosen 
as in the Fig. \protect\ref{compare}. The imaginary parts of Mermin- and the 
VFP-DF are identical.}
\end{figure}
\narrowtext
\twocolumn

\subsection{Comparison of both dielectric function}
\label{comp}

Up to now we have used different kinetic equations leading to two
different dielectric functions. It is of great interest how these
functions are related to each other and whether these
dielectric functions are valid in the storage ring plasma's realm of
temperature, density and friction coefficient.
In Fig.\ref{compare} and Fig. \ref{compare1} are plotted both dielectric functions in dependence on
the frequency for several wave numbers and temperatures.

We see that within the numerical accuracy of the picture no
difference is visible between the dielectric function  of a Fokker Planck
collision integral and the Mermin dielectric function  resulting from the
conserving relaxation time approximation.

Let us inspect now some special limits.
Both dielectric function  fulfills the static limit ($ \omega \rightarrow 0$)
\be \label{r1}
\epsilon(q,0) = 1 + \frac{\kappa^2}{q^2}
\ee
for all $\lambda$ in accordance with the classical Debye-H\"uckel result
for static screening. In the long-wavelength limit $ q \rightarrow 0$
one gets
the Drude formula for both dielectric functions 
\be \label{Drude}
\lim\limits_{\matrix{q\to 0\cr \omega \to \infty}} \epsilon(q,\omega) = 
1 - \frac{\omega_{pl}^2}{\omega(\omega + i \lambda)}.
\ee
For $ \lambda \longrightarrow 0 $ this formula reproduces the RPA behavior.
In the limit of strong friction $\lambda \to \infty$ we get in concur with 
\cite{Park80} and \cite{Lebe95} also (\ref{Drude}).
The long-wavelength and the strong friction limits are identical.

For low temperatures there are differences between the RPA-dielectric function  and the other
correlated dielectric functions. The real parts start in the static limit at the same value as
the RPA-dielectric function  but drops down much more earlier (in Fig.\ref{compare}
one sees 4 magnitudes at one tenth of the inverse Debye length and 2 magnitudes at ten
times the inverse Debye length). There are no zeros in the real part.
Accordingly, the imaginary part is shifted in the same fashion. It is one
magnitude broader than the RPA imaginary part and has only two thirds of
its height.
For temperatures higher than $50K$ the RPA-dielectric function  and the 
Mermin- and VFP dielectric functions become identical.

\subsection{Sum rules}
\label{sumrules}

The most interesting question is whether the dielectric function  fulfills
the sum rules (\ref{fsumrule})-(\ref{s0sumrule}). Due to (\ref{Drude}) all 
presented dielectric functions lead to $ \re\epsilon(\vec q, \omega)
\propto \omega^{-2}$ for large $\omega$. Since poles due to the relaxation 
time occur only in the lower half plane we have
\be
\int\limits_{-\infty}^{\infty}\! \dd\bar\omega \; 
{\epsilon^{-1}(\vec q,\bar\omega) - 1\over
(\bar\omega +i\eta)} = 0
 \ee
from which we see that the dielectric functions fulfill the Kramers-Kronig 
relations
\be\label{kram}
\re \epsilon^{-1}(\vec q, \omega) - 1 = {\cal P} \int_{-\infty}^{\infty}
\frac{2 \, \im \epsilon^{-1}(\vec q,\bar\omega)}{\omega - \bar\omega}
\frac{\dd \bar\omega}{2\pi}
\ee
where the $\cal P$ denotes here the Cauchy principle value.

From (\ref{kram}) we get with (\ref{r1}) in the static limit
just the compressibility sum rule (\ref{f0sumrule}).
The longitudinal f-sum rule (\ref{fsumrule}) follows as well from (\ref{kram}).
To see this we observe that due to time reversibility 
$\epsilon(q,-\omega)=\epsilon^*(q,\omega)$ holds and we can write
\be
\lim\limits_{\omega \to \infty} \re
\epsilon^{-1}(q,\omega) =1+\lim\limits_{\omega \to \infty}{2\over \omega^2
\pi}\int\limits_0^{\infty}
\dd\bar \omega\im \epsilon^{-1}(q,\bar \omega).
\ee
Using (\ref{Drude}) we obtain just the
f-sum rule (\ref{fsumrule}).

Since the same Kramers-Kronig relation (\ref{kram}) holds also for $\epsilon$ 
instead of $\epsilon^{-1}$ we see that the corresponding free sum rules 
(\ref{ssumrule}) and (\ref{s0sumrule}) are also fulfilled.

This completes the proof that both correlated dielectric functions fulfill 
the sum rules. We can state therefore that the dielectric function is properly 
valid in the interesting scope and can be used to describe the phenomena in 
cold and dilute storage ring plasmas. Since the Mermin dielectric function  
is computational much easier to handle, in fact no more circumstances than 
the Lindhard dielectric function, we will use the Mermin dielectric 
function further on.

\section{Application to storage ring plasmas}

We continue now to apply the correlated dielectric function
derived in the last paragraph to typical storage ring plasmas.
Two important quantities we like to discuss here. First the
stopping power of ions in an electron plasma and second the
occurring plasmon excitations.

\subsection{Stopping power}
\label{SP}

The stopping power, i.e. the energy transfer of a particle to a
plasma is given in terms of the dielectric function
by \cite{MR96}
\begin{eqnarray}\label{lb4}
&&{\partial E_a \over \partial t}=-{2 \over \hbar} \int {\dd^3q \over (2 \pi
\hbar )^3}
\; \hbar \omega \; n_B(\hbar \omega) \; {V}_{aa}(q)^2 \; {\rm Im}
\varepsilon^{-1}(q,\hbar \omega).\nonumber\\
&&
\end{eqnarray}
Here $n_b$ denotes the Bose function and $V_{aa}$ the Coulomb
potential of the particle $a$. We observe that the sum
about different plasma species is condensed in the dielectric function.
It is noteworthy to remark that this
result is valid for any arbitrary degeneracy. The derivation presented in
\cite{MR96}
shows that the result (\ref{lb4})
is more general valid than derived earlier
\cite{KS88,BKLMSS901,BKLMSS902}. 
Higher order correlations like vertex corrections can
be incorporated in the dielectric function, such that (\ref{lb4}) remains
valid \cite{MR96}.
This fact is important for dense solid
state plasmas which are used recently for stopping experiments, where the
result (\ref{lb4}) is applicable as well.
A more explicit form can be given by carrying out the angular
integration [$ q = \hbar k $]
\begin{eqnarray}
\label{lb5}
&&{\partial E_a \over \partial t}={2 e_a^2
 \over \pi \varepsilon_0 } {1 \over v(t)}
\int\limits_0^{\infty} \! {\dd k \over k} \int \limits_{-v(t) k
+{\hbar k^2 \over 2 m_a}}^{v(t) k
+{\hbar k^2 \over 2 m_a}} \!\!\!\!\!\!\!\!\! \dd\omega \;
\omega \; n_B(\omega) \; {\rm Im} \varepsilon^{-1}(\hbar k,\omega).\nonumber\\
&&
\end{eqnarray}
Neglecting the quantum effects in (\ref{lb5}) which represent an
internal ionic cut-off due to the thermal De Broglie wavelength
we get the standard result of dielectric theory
\begin{eqnarray}\label{lb6}
{\partial E_a \over \partial t}={2 e_a^2
 \over \pi \varepsilon_0 } {1 \over v(t)}
\int\limits_0^{\infty} \!{\dd k \over k} \int \limits_{0}^{v(t) k} \!\dd\omega \;
\omega  \; {\rm Im} \varepsilon^{-1}(\hbar k,\omega)
\end{eqnarray}
from which all known special cases can be derived \cite{MR96}, among them the
well known Bethe formula. We use (\ref{lb5}) where no artificial cut-off is
needed further on.

In figure \ref{be7_1} we have plotted the stopping power of $^9$Be$^+$ 
calculated with the Mermin and the Lindhard dielectric function. We observe 
that for a weakly 
coupled storage ring plasma with temperature of $6$ K and a density of 
$2.3\, 10^{13}{\rm m}^{-3}$ which corresponds to a nonideality of 
$\Gamma=0.13$ almost no differences are observed between the Mermin and Lindhard 
result. For higher coupling by lower temperature of $1$ K corresponding to 
$\Gamma=0.77$ we see that the Mermin stopping power become smaller than the 
Lindhard result. Since the friction is dependent on the squared of density but 
only on temperature via the Coulomb logarithm we find a stronger dependence 
on the density. This is illustrated in the figures \ref{be7_1}-\ref{be7_3}. 
We see that 
with increasing density the deviations between Mermin and Lindhard results 
become appreciable.

\begin{minipage}[t]{8cm}
\begin{figure}
\centerline{\psfig{figure=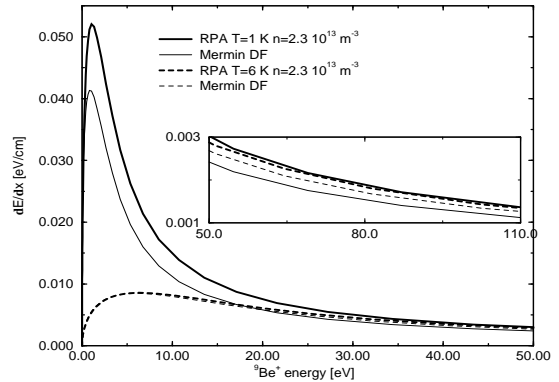,angle=-90,width=8cm,height=6cm}}
\caption{\label{be7_1}
The stopping power of $^9$Be$^+$ ions in an electronic plasma versus 
         ion energy. The classical Lindhard result (thick lines) are compared 
         with the Mermin result (thin lines) for two different temperatures. 
         The plasma parameters are $\Gamma=0.77$ (solid lines) and 
         $\Gamma=0.13$ (dashed lines) respectively.} 
\end{figure}
\end{minipage}

\begin{minipage}[t]{8cm}
\begin{figure}
\centerline{\psfig{figure=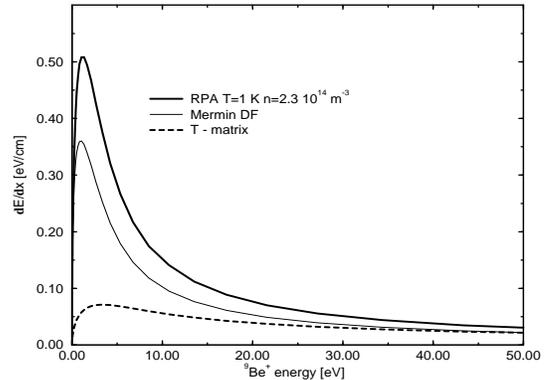,angle=-90,width=8cm,height=6cm}}
\caption{\label{be7_6}
The stopping power of $^9$Be$^+$ ions in an electronic plasma versus 
         ion energy. The classical Lindhard result (thick line) is compared 
         with the Mermin result (thin line) and the T-matrix result (dashed 
         line) of binary collisions (\protect\ref{tmat}). The plasma 
         parameter is $\Gamma=1.65$.}
\end{figure}
\end{minipage}

\begin{minipage}[t]{8cm}
\begin{figure}
\centerline{\psfig{figure=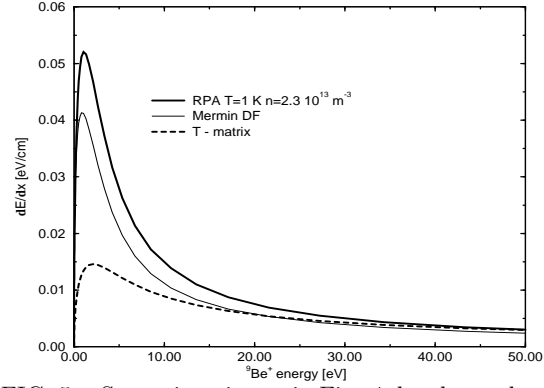,angle=-90,width=8cm,height=6cm}}
\caption{\label{be7_3}
         Same situation as in Fig. \protect\ref{be7_6}, but lower density 
         $ \Gamma = 0.77 $.}
\end{figure}
\end{minipage}
\hfill
\begin{minipage}[t]{8cm}
\begin{figure}
\centerline{\psfig{figure=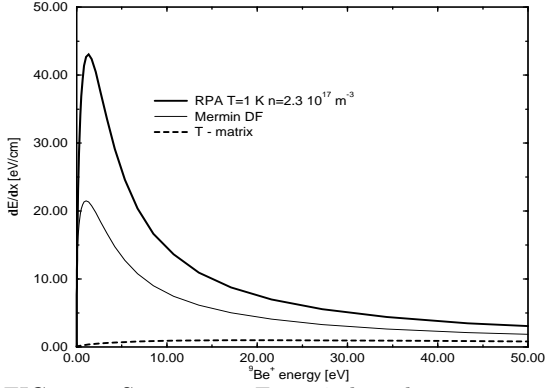,angle=-90,width=8cm,height=6cm}}
\caption{\label{be7_4}
         Same as in Fig. \protect\ref{be7_6}, but density is now 4 magnitudes higher 
         ($ \Gamma = 16.5 $)}
\end{figure}
\end{minipage}

\begin{figure}
\centerline{\psfig{figure=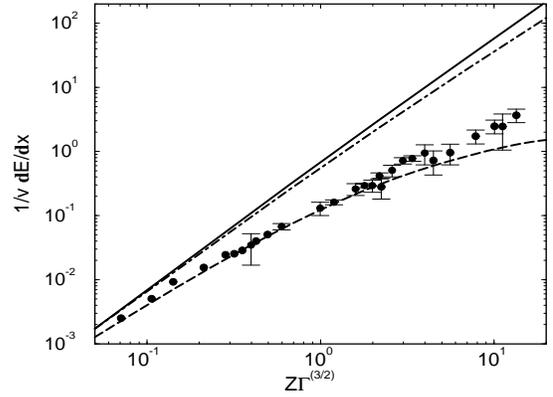,angle=-90,width=8cm,height=6cm,rheight=7cm}}
\caption{\label{Ergebnis}
         The normalized friction coefficient (energy loss) for 
         RPA(solid line), T-matrix result (dashed line) and Mermin DF result
         (dot-dashed line). The  filled circles are simulation results
         \protect\cite{Zwic96} which reproduces experimental data 
         \protect\cite{Wolf94,Wink96-1}.} 
\end{figure}

So far we have generalized the dielectric theory of stopping power by the inclusion
of collisions. It is instructive to compare now the results directly with the 
stopping power in binary collision approximation. In \cite{MR96} the 
following expression for the stopping power was derived from the Boltzmann equation
within T-matrix approximation
\begin{eqnarray}
\label{tmat}
&&{{\partial E \over \partial t}} (v) =\sum\limits_b {n_b v_t \over m_b^2 \sqrt{\pi}}
{e^{-{m_b
v^2 \over 2 k_B T}} \over v} \int\limits_0^{\infty}\! \dd p \; p^2
\sigma^t_{ab}(p) \nonumber\\
&&\left [ a \, {\rm cosh} \, a-(1+ {p^2 (1+{m_b \over m_a})
\over
m_a k_B T}) {\rm sinh} \, a \right ] e^{-{p^2 \over 2 m_b k_B T} (1+{m_b
\over m_a})^2}\nonumber\\
&&
\end{eqnarray}
with the thermal velocity $v_t^2=2 k_B T/m_b$, the abbreviation
$a={v p \over k_B T} (1+{m_b \over m_a})$ and the quantum mechanical
transport cross section 
\begin{eqnarray}\label{cross}
\sigma^t(p)=\int\! \dd\Omega \, (1-{\rm cos}\theta) {\dd \sigma \over
\dd \Omega}.
\end{eqnarray}
In \cite{RR89} a fit formula is given which
subsumed the numerical results for the transport cross
section for a plasma with charge $Z=1$. In figures \ref{be7_1} - \ref{be7_3} 
we compare the results for the dielectric theory of stopping power with and 
without collisional contributions with the pure two- particle collision result
of (\ref{tmat}). We see that the two-particle collision expression is 
significantly smaller than the dielectric theory. For very strong coupling 
in figure \ref{be7_4} we see even a vanishing contribution of the latter 
one indicating that the two-particle collisions do not contribute 
any more but the energy transfer is mainly caused by collective spreading.
In Fig.\ref{Ergebnis} we represent the reduced energy loss [$\lambda_{\rm l}=e^2/(12 \pi \epsilon_0 T)$]
\be
\frac{\lambda_{\rm l} v_{th}}{k_B T}\frac{1}{v} {d E \over dx}
\ee
versus the coupling parameter $ Z \Gamma^{3/2}$.

The dependence of the normalized energy loss from the coupling parameter 
is weaker in the Mermin case than in the RPA case but distinct from the
numerical simulations. Nevertheless the involving of collisions modifies
the stopping power in the right direction. The best description is still given by the T-matrix result (\ref{tmat}).

\subsection{Plasmons}
\label{Schottky}

The ion beam current $ j(t)$ is a fluctuating quantity due to its granular 
(ionic) 
structure. Detecting the mirror charge on the vacuum chamber of the ring
and Fourier transforming (frequency analyzing) one obtains the 
Schottky-signal. It is primarily used for analyzing the beam's velocity 
distribution and hence the longitudinal temperature, but also important for 
measuring the particle number or the revolution frequency of the beam. 
It is related to the dynamical structure factor $ S(\vec q,\omega) $ by the 
equation \cite{Avil93}
\be
\left< \left| J(q,\omega) \right|^{2} \right> \sim S(q,\omega) \;,
\ee
where the brackets indicate the thermal averaging.
The well-known fluctuation-dissipation theorem \cite{Kubo57} connects
the imaginary part of the response function
$ \mbox{Im} \epsilon(\vec q,\omega)$ and the dynamical structure factor
\be
\label{FlukDissTh}
\nonumber
S(\vec{q},\omega) =
-\frac{k_{B}T}{\omega V_C (\vec{q})}\im\epsilon^{-1}(\vec{q},\omega)
\ee
with the Coulomb potential $V_C$.
In dense ion beams (e.g. the C$^{6+}$ LIR experiment) one observes a
double peaked Schottky spectrum. The two peaks commonly identified as
plasma waves propagating in two directions around the storage ring.
A frequency analysis of this beam current shows the propagating waves
clearly as peaks in the spectra which are theoretically well described in
\cite{Park80,Lebe95}. The identity of Mermin and VFP-dielectric function
we use now to compute the the Schottky noise much easier within the Mermin
DF.

For numerical calculation one has to modify the plasma frequency which differs
from the one of an isotropic plasma. For a plasma in a conducting tube we
have 
\be
\label{Omega}
\tilde\omega_{pl}^2 = \frac{NZ^2 e^2 M}{2\pi R_0^3 m \epsilon_0 \gamma}
         \left( \ln{\frac{r_C}{r_B}} + \frac{1}{2}\right)
\ee
Here denotes $N$ the particles number, $ r_C $ the radius of the beam chamber,
$ r_B $ the radius of the beam propagating with the velocity 
$v_0 = \gamma c$ (speed of light), $ 2\pi R_0 $ is the circumference of the 
ring. $M$ is the number of plasma waves fitting in the ring, the so-called
harmonic wavenumber.

The equation for the plasma frequency follows straightforward from inserting
the beam pipe's impedance $ {\cal Z} $ via Ohm's law 
$ U = e^2 v_0 {\cal Z}/R_0 $ in the
external field of the Vlasov equation. Usually $ {\cal Z} $ is given by
\be
{\cal Z} = \frac{M}{v_0 \gamma^2}
    \left(\ln\left(\frac{r_C}{r_B}\right) + \frac{1}{2}\right) \;.
\ee
We assume for simplicity the non-relativistic case which is valid in the
TSR experiments ($^9$Be$^{+}$: $0.04c$, $^{12}$C$^{6+}$: $ 0.15c $) with
$ \gamma \approx 1$.

\begin{figure}
\centerline{\psfig{figure=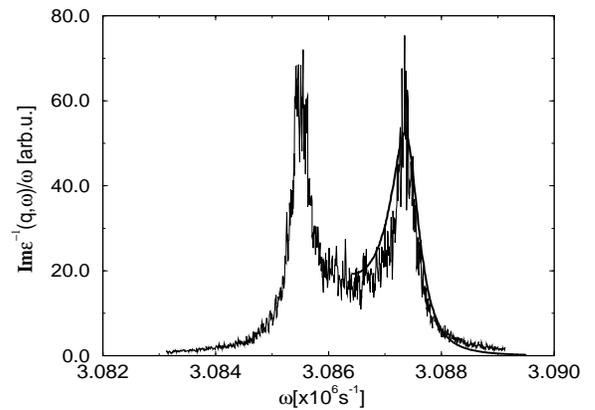,angle=-90,width=8cm,height=6.0cm,rheight=6.5cm}}
\caption{\label{Schottkyeps}
         The Schottky spectra of a dense carbonium beam and the corresponding
         theoretical prediction. ($ n = 8.3\E{13}$m$^{-3}$, $\Tl = 11000$K).
         Data are taken from \protect\cite{Tetz97}.}
\end{figure}

The wavelength $q$ is not a
continuous variable but assumes only discrete values $q=M/2\pi R_0 $ .
The fraction $\kappa^2/q^2 $ can now expressed by
\be
\frac{\kappa^2}{q^2} = 2\frac{\tilde\omega_{pl}^2}{\delta \omega^2}
\ee
with the thermal frequency
\be
\label{delom}
\delta \omega^2 = \frac{2 \kBT}{R_0^2 m} \;.
\ee
Inserting this parameters into the VFP dielectric function  one obtains
\be
&&\epsilon_{\rm VFP}(M,\omega) = \nonumber\\
&&1 + 2
  \left(\frac{\tilde\omega_{pl}}{ \delta\omega} \right)^2\!\!
  _1F_1 \left[1, 1 + \frac{M^2 \delta\omega^2}{2\lambda^2} -
\frac{i\omega}{\lambda},\frac{M^2 \delta\omega^2}{2\lambda^2} \right] \;.
\ee
This is the well known standard permittivity in the Schottky noise theory.
Since the Mermin DF and the VFP-DF are identical we use practically now the
easier Mermin DF.
We modify equation (\ref{MerminDF}) according to the parameters 
(\ref{Omega})-(\ref{delom})
\be
\epsilon_{\rm M} \left( M,\omega+ i\lambda \right) &=& 1 + \frac{
  \left(1+ \frac{\lambda}{\omega}\right)
  (\epsilon(M,\omega+i\lambda) - 1) }
{ \displaystyle
1 + \frac{i\lambda}{\omega} \frac{ \epsilon( M,\omega+i\lambda) - 1}
                                { \epsilon( M,0) - 1}}) \\
\ee
with (\ref{RPADF}) for $ \epsilon(M,\omega + i\lambda) $ and
\be
\epsilon_{\rm M}(M,0) &=& 1 + 2\frac{\tilde\omega_{pl}^2}{\delta \omega^2} \;.
\ee
In the next step we insert a relaxation time considering the  
anisotropy in the thermal velocities $ v_{th} $ \cite{Sore87}
\be
\lambda = 4\pi (Ze)^4 \frac{n}{\epsilon^2 m^2 v_{th,\perp} v_{th,\parallel}^2}
\Lambda \,.
\ee
Here $\Lambda $ again denotes the Coulomb logarithm. We have used the 
modified Mermin-dielectric function  for calculating the plasmonic excitation
for a $^{12}$C$^{6+}$ beam at an energy of $ 73.3$MeV and a revolution 
frequency of $ v_0/R_0 = 617$kHz ($ M = 5 $). In \ref{Schottkyeps} the 
expression $\im\epsilon(M, \omega)/\omega$ is compared with the 
Schottky measurement. We see that the modified Mermin-dielectric function  
fits the Schottky spectra satisfactorily.

\section{summary}

In this paper we have described two dielectric functions including collisions.
After numerical inspection we have shown that the DF obtained from VFP 
equation is identical with the Mermin DF. Modifying the dielectric function  
for storage ring purposes (modified plasma frequency, discrete 
wavenumbers/harmonics) we have derived the standard dielectric function 
for Schottky noise prediction for a three dimensional plasma beam.  Because of
the identity of both dielectric functions one can use a Mermin dielectric 
function  for Schottky noise description, too. 
The second goal was
a better description of the stopping power acting on an ion beam in the
cooler's electron gas. Here we like to state, that including the collisions
leads to a lower friction force than in the RPA-predictions but obvious 
overestimates the friction force compared with simulated and experimental
results.

Further improvement has to be done to consider strong coupling effects in the 
relaxation time by using a matching Coulomb logarithm. Efforts have also to be 
done to include the magnetic field in the cooler and the anisotropic Maxwell 
distributions in the dielectric function.


\begin{appendix}
\section{ Solution of the Vlasov-Fokker-Planck equation}

We following the the main ideas of \cite{Resi96}.
At first we introduce reduced variables
\be
\vec{u} = \sqrt{ \frac{m}{2 \kBT}} \vec{v} \qquad
\vec{k} = \frac{\vec q}{\lambda}\sqrt{\frac{\kBT}{2m}} \;.
\ee
After linearization the VFP-equation (\ref{VFPE}) we arrive at
\be
\nonumber
& & \frac{\partial}{\partial t} \delta f(\vec k, \vec u, t)
+ 2i\lambda \vec u \vec k\; \delta\! f(\vec k, \vec u, t)
- \frac{2i\lambda}{m} \vec{k} U(\vec k,t)
\frac{\partial f_0(\vec{u})}{\partial \vec u}  \\
&& = \lambda \frac{\partial}{\partial \vec u} \left( \frac{1}{2}
\frac{\partial}{\partial \vec u} + \vec u \right) \; .
\ee
For the Fokker-Planck term including the second term of the left side (the so-called
inhomogeneous Fokker-Planck operator) one can consider
the eigenvalue and eigenfunctions in three dimensions
\be
&&\left[ -2i\lambda\vec u \vec k +
\lambda \frac{\partial}{\partial \vec u}\left(
\frac{1}{2}\frac{\partial}{\partial \vec u} + \vec u \right) \right]
\psi_{\nxyz}(\vec k, \vec u) \\
\nonumber
&& = \Lambda_{\nxyz}(\vec k) \psi_{\nxyz}(\vec k, \vec u) \;.
\ee
The eigenfunctions take the form after a coordinate transformation
$ z = u + 2ik $
\be
\psi_{\nxyz} &=& \sqrt{ \frac{1}{n_x! n_y! n_z! 2^{n_x+n_y+n_z}}
\left(\frac{m}{2\pi \kBT}\right)^3 }\nonumber\\
&\times&
{\rm e}^{k^2} {\rm e}^{-(\vec{z} - i\vec{k})^2} 
\Herm_{n_x}[z_x] \Herm_{n_y}[z_y] \Herm_{n_z}[z_z] \;.
\ee
One can identifying the functions $\Herm_n$ as Hermite-Polynoms.
The lowest eigenfunction for $ n_x=n_y=n_z=0 $ is the Maxwell-distribution.
The eigenvalues are
\be
\Lambda_{n_{x}n_{y}n_{z}} = -\lambda (n_{x}+n_{y}+n_{z} + 2 k^2) \;.
\ee
In the next step we insert for the right side the eigenvalues and
eigenfunctions of the inhomogeneous Fokker-Planck operator
\be
& & \frac{\partial}{\partial t} \delta\! f(\vec k, \vec u, t)
- \frac{2i\lambda}{m} \vec{k} U(\vec k,t)
\frac{\partial f_0(\vec{u})}{\partial \vec u}   \\
\nonumber
&=& \lambda \sum_{\nxyz} \mu_{\nxyz} \psi_{\nxyz} \\
\nonumber
\mbox{with}
& & \mu_n(\vec k) = \frac{ \Lambda_n(\vec q)}{\lambda}
\ee
and expand the distribution function $ \frac{\partial}{\partial \vec{u}} f_0 $
and the distortion $ \delta\! f(\vec{u},\vec{k},t) $ in a sum of
eigenfunctions of the Fokker-Planck operator
\be
\vec{k} \frac{\partial}{\partial \vec{u}} f_0(\vec{u})
&=& \vec{k}\sum_{\nxyz}^{\infty} \vec{a}_{\nxyz} \psi_{\nxyz}(\vec{u},\vec{k})
\nonumber\\
\delta\! f(\vec{u},\vec{k},t) &=& \sum_{\nxyz} c_{\nxyz} \psi_{\nxyz} \;.
\ee
Performing a Fourier transformation $ t \rightarrow \omega $ we arrive at a
solution for the coefficients
\be
c_{\nxyz} = \frac{ \frac{i\lambda}{\kBT} \vec{k}\vec{a}_n U(\vec{k},\omega)}
{-i\omega - \lambda (n_x+n_y+n_z + 2k^2)} \;.
\ee
Remembering (\ref{deln-Pi}) we obtain
\be
\delta\! n(\vec{k},\omega) &=& \sum_{\nxyz} c_{\nxyz}
\int\limits_{Vol} \psi_{\nxyz} \dd^3 u \\
\nonumber
&=& U(\vec{k},\omega) \Pi_{\rm VFP}(\vec{k},\omega).
\ee
Hence we get for the polarizability
\be
&&\Pi_{\rm VFP}(k,\omega) =  \frac{\lambda}{2\pi\kBT} \exp[k^2]
\sum_{\nxyz} i^{2(n_x+n_y+n_z)} \nonumber
\\
&\times& \frac{\Gamma\left[ \frac{1}{2} + n_x \right]
          \Gamma\left[ \frac{1}{2} + n_y \right]
          \Gamma\left[ \frac{1}{2} + n_z \right]
          \Gamma\left[ \frac{1}{2} + n_x + n_y \right] }
         {n_x! n_y! n_z! (n_x + n_y)! (n_x + n_y + n_z)!} \nonumber
\\
&\times& \frac{ (n_x+n_y+n_z + 2k^2) (\sqrt{2} k)^{2(n_x+n_y+n_z)}}
         {-i\omega - \lambda (n_x + n_y + n_z + 2 k^2)} \;.
\ee
Using the relations
\be
\nonumber
 &\sum_{n_x,n_y,n_z}& \!
  \frac{\Gamma\left[ \frac{1}{2} + n_x \right]
        \Gamma\left[ \frac{1}{2} + n_y \right]
        \Gamma\left[ \frac{1}{2} + n_z \right]
        \Gamma\left[ \frac{1}{2} + n_x + n_y \right] }
     {n_x! n_y! n_z! (n_x + n_y)! (n_x + n_y + n_z)!}  \\
&=& \sum_{m} \frac{\pi^2}{m!} \;,
\ee
and
\be
\sum_n^{\infty} \frac{(-x)^n}{n! (\kappa + n)}
= \frac{\exp[-x]}{\kappa} \left. _1F_1[1,1+\kappa,x] \right. 
\ee
we arrive finally at
\be
&&\Pi_{\rm VFP}(k,\omega) = \\
\nonumber
&&\frac{\pi}{2 \kBT} \left( 1 +
\frac{i \omega}{2\lambda k^2 - i \omega}
\left. _1F_1\right.
\left[1, 1+2k^2 - i\frac{\omega}{\lambda}; 2k^2 \right] \right) \;.
\ee

The dielectric function related to this polarizability was discussed in chapter
\ref{VFPDF_ch}.

\end{appendix}

\acknowledgements
The authors acknowledge stimulating discussions with G. R\"opke.
This work was supported from the BMBF (Germany) under
contract Nr.~06R0884 and the Max-Planck-Society.

\end{document}